\documentclass[final,3p,fleqn]{elsarticle}

\usepackage{epsfig}
\usepackage{epstopdf}

\usepackage{amssymb}
\usepackage{amsmath}
\usepackage{graphics}
\usepackage[labelformat=empty]{caption}

\usepackage{amsthm}
\usepackage{natbib}

\usepackage{lineno}

\journal{Arxiv}

\begin{document}
\begin{frontmatter}



\title{\bf{Self similar flow under the action of monochromatic radiation behind a cylindrical shock wave in a self-gravitating, rotating  axisymmetric dusty gas}}

\author{Ruchi Bajargaan}

\ead{duruchi11@gmail.com, +91-8130576177}

\author{Arvind Patel*}
\cortext[mycorrespondingauthor]{Corresponding author.}
\ead{apatel@maths.du.ac.in, +91-9310568439}


\address{Department of Mathematics, University of Delhi, Delhi 110 007, India}

\begin{abstract}
The propagation of a cylindrical shock wave in a self-gravitating, rotating axisymmetric dusty gas under the action of monochromatic radiation with a constant intensity per unit area, which has variable azimuthal and axial components of fluid velocity, is investigated. The gas is assumed to be grey and opaque, and the shock is assumed to be transparent. The dusty gas is considered as a mixture of non-ideal gas and small solid particles, in which solid particles are continuously distributed. To obtain some essential features of shock propagation, small solid particles are considered as a pseudo-fluid, and it is assumed that the equilibrium flow condition is maintained in the entire flow-field. Similarity solutions are obtained as well as the effects of the variation of the radiation parameter, the gravitation parameter, the non-idealness parameter of the gas, the mass concentration of solid particles in the mixture, the ratio of the density of solid particles to the initial density of the gas are worked out in detail. The similarity solution exists under the constant initial angular velocity, and the shock strength is independent from the radiation parameter and the gravitation parameter. It is found that radiation parameter dominates the effect of dusty gas parameters on the variation of radiation heat flux. The total energy of the flow-field behind the shock front is not constant but varies as fourth power of the shock radius.
\end{abstract}


\end{frontmatter}
\section{Introduction}

Shock processes can normally take place in a variety of astrophysical situations such as supernova explosions, photo ionized gas, stellar winds, collision between high velocity clumps of interstellar gas, etc. Shock phenomena , for example, a global shock resulting from a stellar pulsation, supernova explosion passing outward through a stellar envelope, or may be a shock arising from a point source such as a man-made explosion in Earth's atmosphere or an impulsive flare in Sun's atmosphere, have immense importance in astrophysics and space sciences. Shocks are ubiquitous through out the observe universe and play a essential role in the transportation of energy into the interstellar medium, setting in motion processes observed in nebulae that ultimately can lead to the creation of new stars. Shock waves are common in the interstellar medium due to a great variety of supersonic motions and energetic events, such as supernova explosions, cloud-cloud collision, bipolar outflow from young protostellar objects, powerful mass losses by massive stars in a late stage of their evolution (stellar winds), supernova explosions, central part of star burst galaxies etc. Shock waves are also connected with spiral density waves, radio galaxies, and quasars. Same phenomena also takes place in laboratory situations, for example, when a piston is driven rapidly into a tube of gas (a shock tube), when a projectile or aircraft moves supersonically through the atmosphere, in the blast wave produced by a strong explosion, or when rapidly owing gas encounters a constriction in a flow channel or runs into a wall.

The study of shock waves in a mixture of small solid particles and a gas is of great importance due to its applications in lunar ash flow, coal-mine blast, nozzle flow, bomb blast, metallized propellant rocket blast, underground, volcanic, and cosmic explosions, supersonic flight in polluted air, collision of coma with a planet, description of star formation, particle acceleration in shocks, formation of dusty crystals, and many other engineering problems \cite{pai, higa, miura, miu, gret1}. Recently, applications of dusty-gas flow studies to environmental and industrial issues have drawn attention. When a moving shock wave hits a two-phase medium of gas and particles then a flow-field develops which has a close practical relation to industrial applications (solid rocket engine in which aluminium particles are used to decrease the vibration because of instability) as well as industrial accidents such as explosions in coal mines and grain elevators \cite{park}.  Miura and Glass \cite{miura} have obtained an analytical solution of a planar shock wave in dusty gas with constant velocities. As the volume occupied by solid particles mixed into a perfect gas is negligibly small, dust virtually has a mass fraction rather than a volume fraction. The results of \cite{miura} reflect the influence of the additional inertia of dust upon shock propagation. Pai et al. \cite{pai} have generalized the well-known solution of a strong explosion because of an instantaneous release of energy in a gas to the case of two-phase flow (i.e. mixture of small solid particles and perfect gas) and obtained the key effects due to presence of dust particles on such a strong shock wave. As a nonzero volume fraction of solid particles in the mixture was assumed, the results reflect the influence of both the decrease in the mixture compressibility and the increase in the mixture inertia on shock propagation. Higashino and Suzuki \cite{higa} have studied the line source explosions in a dusty gas with the assumptions of velocity and temperature equilibrium. Steiner and Hirschler \cite{steiner} have obtained analytic solutions for the one-dimensional unsteady self-similar flow of a dusty gas between a strong shock and a moving piston behind it. The results of \cite{steiner} reflect that the dust's inertia and its solid phase behaviour, strongly influence the wave propagation.

In recent years, considerable attention has been given to the study of the interaction between gasdynamics and radiation. In gasdynamics, if the radiation effects are taken into account, the basic non-linear equations become complicated, and therefore it is necessary to establish such approximations that are physically accurate and can bear significant simplification. The problems of the interaction of radiation with gas dynamics have been studied by several authors by using the theory of dimensionality which was developed by Sedov \cite{Sedov}. Marshak \cite{marshak} has obtained similarity solutions of the radiation hydrodynamic equations for particular cases when there is plane symmetry, and radiation pressure and energy are negligible, although flux is important. Marshak \cite{marshak} considered the cases of (1) constant density, (2) constant pressure, and (3) power law time dependence of temperature. Elliott \cite{Elliott} discussed the conditions leading to self-similarity with a specified functional form of the mean-free path of radiation. Wang \cite{wang}, Helliwell \cite{helli} and Nicastro \cite{nic} considered the problems of stationary or moving radiating walls generating shock at the head of self-similar flow-fields. Ray and Bhowmick \cite{ray} obtained the self-similar solution for the central explosions in stars with radiation flux by taking the shock to be isothermal and transparent. Khudyakov \cite{khud} considered the self similar problem of the motion of a gas under the monochromatic radiation. Zheltukhin \cite{zhel} developed a family of exact solutions of one dimensional motion (plane, cylindrical, or spherical symmetry) of a perfect gas by taking the absorption of monochromatic radiation. Verma, Srivastava and Khan \cite{ve} considered homothermal magnetogasdynamic shock waves caused by instantaneous monochromatic radiation. Nath and Thakar, and Nath \cite{na, na1} studied the propagation of cylindrical shock waves in rotating or non-rotating perfect gas under the action of monochromatic radiation. Shinde \cite{sh} obtained the similarity solution of the propagation of magnetogasdynamic cylindrical shock waves in a non-uniform, rotating perfect gas under the action of monochromatic radiation and gravitation. Vishwakarma and pandey \cite{vis} obtained the similarity solution for one-dimensional flow under the action of monochromatic radiation behind a cylindrical magnetogasdynamic shock wave propagating in a non-ideal gas. Nath, sahu and Dutta \cite{na3} obtained similarity solution of magnetohydrodynamic cylindrical shock wave in a non-uniform rotating non-ideal gas under the action of monochromatic radiation. Nath and Sahu \cite{na4} obtained similarity solution of cylindrical shock wave in a rotational axisymmetric non-ideal gas under the action of monochromatic radiation. Nath and Sahu \cite{na6} obtained similarity solution of cylindrical shock wave in a non-ideal dusty gas under the action of monochromatic radiation. Sahu \cite{sahu} has investigated the propagation of a cylindrical shock wave in a rotational axisymmetric non-ideal gas under the action of monochromatic radiation with increasing energy by assuming variable azimuthal and axial fluid velocity.

The investigation of the most important celestial phenomena must be centred on the problems of motion of gaseous masses with shock waves in a gravitational field. The gravitational forces have considerable effects on several astrophysical problems. The unsteady motion of a large mass of gas followed by a sudden release of energy results in flare ups in novae and supernovae. A qualitative behaviour of the gaseous mass may be discussed with the help of the equations of motion and equilibrium taking gravitational forces into account. Carrus et al. \cite{carrus} have obtained the similarity solutions of the propagation of shock waves in a gas under the gravitational attraction of a central body of fixed mass (Roche model) by numerical method. Rogers \cite{rogers} has discussed a method for obtaining an analytical solution of the same problem. Patel \cite{patel} has obtained self-similar solutions for the one-dimensional unsteady adiabatic flow of a dusty gas behind spherical shock wave under the gravitational field. Recently, the study of self-similar solution of a shock wave in a perfect gas, or non-ideal gas, or dusty gas with the gravitational field has been done by many authors \cite{ss, ruchi, baj}.

In all of the above mentioned works, the effect of gravitational field on the shock propagation in rotating axisymmetric non-ideal dusty gas under the action of monochromatic radiation is not studied. The effects of small solid particles under the action of monochromatic radiation and gravitational field are not taken into account by any of the authors. In the present work, we generalize the solution of Nath and Thakar in perfect gas to the case of a dusty gas (a mixture of non-ideal gas and small solid particles) and also by taking into account the gravitational field, rotation of the medium and the components of the vorticity vector with a slightly different transformation.

The purpose of this study is to obtain self similar solutions for the flow behind a cylindrical shock wave propagating in a rotating axisymmetric non-ideal dusty gas under the action of monochromatic radiation and gravitational field, which contain variable axial and azimuthal fluid velocities. In the ambient medium, the components of fluid velocity are taken to be varying and obeying the power laws. The radiation flux with constant intensity $j_0$ is assumed to move in the opposite direction to shock wave propagation. It is assumed here that the gas itself does not radiate and the radiation is absorbed only behind the shock wave. The gas is assumed to be gray and opaque and shock to be transparent. The shock is assumed to be propagating in a conducting medium at rest. In order to obtain some essential features of shock propagation, small solid particles are considered as a pseudo-fluid, and the mixture at temperature and velocity equilibrium with a constant ratio of specific heats. The thermal conductivity and viscous stress of the mixture of small solid particles and the gas are assumed to be negligible. The effects of change in the variation of non-idealness parameter of the gas, the ratio of the density of solid particles to the initial density of the gas, the mass concentration of solid particles in the mixture, the radiation parameter, and the gravitation parameter are investigated.

\section{Equations of motion and boundary conditions}

The fundamental equations for one-dimensional unsteady adiabatic flow behind a shock wave in a self gravitating, rotating axisymmetric dusty gas (mixture of a non-ideal gas and small solid particles) under the action of monochromatic radiation, neglecting heat-conduction, viscosity, and radiation of the medium, can be presented in the Eulerian coordinates in the following form \cite{pai,  khud, na1, ruchi, baj, levin, zedan}:

\begin{align}
&{\frac{\partial \rho}{\partial t}}+u{\frac{\partial \rho}{\partial r}}+\rho {\frac{\partial u}{\partial r}}+{\frac{ u\rho}{r}} = 0,\label{1.1}\\
&\frac{\partial u}{\partial t}+u\frac{\partial u}{\partial r}+\frac{1}{\rho}\frac{\partial p}{\partial r}+\frac{\bar{G}m}{r}-\frac{v^2}{r} = 0,\label{1.2}\\
&\frac{\partial v}{\partial t}+u\frac{\partial v}{\partial r}+\frac{uv}{r}=0,\label{1.3}\\
&\frac{\partial w}{\partial t}+u \frac{\partial w}{\partial r}=0, \label{1.4}\\
&\frac{\partial m}{\partial r} = 2\pi \rho r,\label{1.5}\\
&\frac{\partial U_{m}}{\partial t}+u\frac{\partial U_{m}}{\partial r}-\frac{p}{\rho^{2}}\biggl(\frac{\partial \rho}{\partial t}+u\frac{\partial \rho}{\partial r}\biggr)=\frac{1}{\rho r}\frac{\partial (F {r})}{\partial r},\label{1.6}\\
&\frac{\partial F}{\partial r}=K F, \label{1.7}
\end{align}

where $r$ and $t$ are the independent space and time coordinates, $u$, $v$ and $w$ are the radial, azimuthal and axial components of the fluid velocity $\vec{q}$ in the cylindrical coordinates $(r, \theta, z)$, $F$ is the flux of monochromatic radiation per unit area at a radial distance $r$ and time $t$, $K$ is the absorption coefficient, $\bar{G}$ is the gravitational constant, and $p$, $\rho$, $m$ and $U_m$ are the pressure, density, total mass per unit volume and internal energy per unit mass of the mixture.

Also,
\begin{equation}\label{1.8}
v=Ar,
\end{equation}
where `$A$' is the angular velocity of the medium at radial distance $r$ from the axis of symmetry. In this case the vorticity vector $\zeta=\frac{1}{2}curl\vec{q}$, has the following components
\begin{equation}\label{1.9}
{\zeta}_r=0,\quad {\zeta}_\theta=-\frac{1}{2}\frac{\partial w}{\partial r}, \quad {\zeta}_z=\frac{1}{2r}\frac{\partial}{\partial r}(rv).
\end{equation}

We consider the medium to be dusty gas i.e. mixture of a non-ideal gas and small solid particles. The equation of state of the non-ideal gas in the mixture is taken as
\begin{equation}\label{1.10}
p_g=R^{*} {{\rho}_g} (1+b {\rho}_g) T, 
\end{equation}
 
 where $p_g$ and ${\rho}_g$ are the partial pressure and density of the gas in the mixture, $R^{*}$ is the specific gas constant, $T$ is the temperature of the gas (and of the solid particles because the equilibrium flow condition is maintained), and `$b$' is the internal volume of the molecules of the gas. 
 
 The specific volume of the solid particles is assumed to remain unchanged without being affected by variations in temperature and pressure. Therefore, the equation of state of the solid particles in the mixture can be expressed as
 \begin{equation}\label{1.11}
{\rho}_{sp}= \text{constant},
 \end{equation}
 
 where ${\rho}_{sp}$ is the specific density of the solid particles.
 
 The equation of state of the mixture of a non-ideal gas and small solid particles can be written as given below \cite{vn, pa}
 \begin{equation}\label{1.12}
 p= \frac{1-K_p}{1-Z}[1+b\rho (1-K_p)]\rho R^* T.
\end{equation}

Here $K_p=m_{sp}/m$ is the mass fraction (concentration) and $Z=V_{sp}/V_m$ is the volume fraction of the solid particles in the mixture, where $m_{sp}$ and $V_{sp}$ are the total mass and volume of the solid particles and $V_m$ is the total volume of the mixture.

The internal energy per unit mass of the mixture can be written as
\begin{equation}\label{1.13}
U_m=\frac{p(1-Z)}{(\Gamma-1)\rho[1+b\rho (1-K_p)]},
\end{equation}

where $\Gamma$ is the ratio of the specific heats of the mixture.

The absorption coefficient $K$ is considered to vary as \cite{khud, na1, na2}
\begin{equation}\label{1.14}
K=K_0 {\rho}^{\alpha} p^{\delta} F^q r^s t^l,
\end{equation}

where $K_0$ (the coefficient of radiation absorption ahead of the shock wave front) is a dimensional constant, and the exponents $\alpha$, $\delta$, $q$, $s$ and $l$ are rational numbers. 

We assume that a diverging cylindrical shock wave is propagating in the dusty gas with a constant density. Therefore, the following equalities are valid for the flow variables immediately ahead of the shock front:
 \begin{align}
 & u=u_a=0,\label{1.15}\\
 & \rho={\rho}_a=\text{constant},\label{1.16}\\
 & v=v_a=v^{*} R^{\lambda},\label{1.17}\\
 & w=w_a=w^{*} R^{\sigma},\label{1.18}
 \end{align}

where $v^{*}$, $w^{*}$, $\lambda$ and $\sigma$ are constants, $R$ is the shock radius, and the subscript `$a$' refers to the conditions immediately ahead of the shock front.

In the undisturbed state of the gas, the equation (\ref{1.2}) and (\ref{1.3}) gives
 \begin{align}
 & m_a=\pi {\rho}_a R^{2},\label{1.19}\\
 & p_a=\frac{-\bar{G} \pi {{\rho}_a}^2 R^2}{2}+\frac{{\rho}_a {v^{*}}^2 R^{2\lambda}}{2\lambda}.\label{1.20}
 \end{align}
 
The components of the vorticity vector ahead of the shock vary as
\begin{equation}\label{1.21}
{\zeta}_{r_a}=0, \quad {\zeta}_{\theta_a}=-\frac{w^{*}\sigma}{2} R^{\sigma-1}, \quad {\zeta}_{z_a}=\frac{(1+\lambda)v^{*}}{2}R^{\lambda-1}.
 \end{equation}
 
 From equations (\ref{1.8}) and (\ref{1.17}), the initial angular velocity of the ambient medium vary as
\begin{equation}\label{1.22}
A_a=v^* R^{\lambda-1}.
\end{equation}
  
  The jump conditions at the shock propagating into non-ideal dusty gas which is transparent for the radiation flux, are given by the conservation of mass, momentum and energy across the shock
  \begin{align}
{\rho}_a \dot{R}&={\rho}_n(\dot{R}-{u_n}),\label{1.231}\\
p_a+ {\rho}_a {\dot{R}}^2&=p_n+ {\rho}_n(\dot{R}-u_n)^2,\label{1.232}\\
U_{m_a}+ \frac{p_a}{\rho_a}+\frac{{\dot{R}}^2}{2}&=U_{m_n}+\frac{p_n}{\rho_n}+\frac{(\dot{R}-u_n)^2}{2},\label{1.233}\\
\frac{Z_a}{\rho_a}&=\frac{Z_n}{\rho_n},\label{1.234}\\
F_a&=F_n,\label{1.235}\\
m_a&=m_n,\label{1.236}\\
v_a&=v_n,\label{1.237}\\
w_a&=w_n,\label{1.238}
\end{align}
where the subscript `$n$' denotes the conditions immediately behind the shock front, $\dot{R}(=\frac{dR}{dt})$ denotes the velocity of shock front.
 
 From equation (\ref{1.15}), the flow variables behind the shock front are given by 
\begin{align}
&{\rho}_n= \frac{{\rho}_a}{\beta}, \label{1.241}\\
&u_n=(1-\beta)\dot{R}, \label{1.242}\\
&p_n={\rho}_a {\dot{R}}^2 \biggl[(1-\beta)+\frac{1}{\gamma M^2}\biggr],\label{1.243}\\
&Z_n=\frac{Z_a}{\beta},\label{1.244}\\
&F_n=F_a,\label{1.245} \\
&m_n=m_a,\label{1.246}\\
&v_n=v_a,\label{1.247}\\
&w_n=w_a,\label{1.248} 
\end{align}

where $M^2=(\frac{{\rho}_a {\dot{R}}^2}{\gamma p_a})^{\frac{1}{2}}$ is the shock-Mach number referred to the frozen  speed of sound $(\frac{\gamma p_a}{\rho_a})^{\frac{1}{2}}$. The quantity $\beta\; (0<\beta<1)$ is obtained by the relation
\begin{align}
&{\beta}^2(\frac{\Gamma+1}{2})-\beta[\frac{\Gamma}{\gamma M^2}+(\frac{\Gamma-1}{2})\{1+\bar{b}(1-K_p)\}+Z_a-(\Gamma-1)\bar{b}(1-K_p)]-(\frac{\Gamma-1}{2})\bar{b}(1-K_p)\nonumber\\
&{}-\frac{(\Gamma-Z_a)\bar{b}(1-K_p)+(\Gamma-1){\bar{b}}^2(1-K_p)^2}{\gamma M^2\{1+\bar{b}(1-K_p)\}}=0,\label{1.25}
\end{align}
where $Z_a$ is the initial volume fraction and $\bar{b}=b{\rho}_a$ is the non-idealness parameter of the gas. Equation (\ref{1.25}) gives two different values of $\beta$ for all the values of parameters $\Gamma$, $\gamma$, $M$, $\bar{b}$, $K_p$ and $Z_a$, out of which only one value lies in the required range $0<\beta<1$, i.e. only one value of $\beta$ satisfies the physical limit of the considered problem.

The jump conditions for the components of vorticity vector across the shock front are given as 
\begin{equation}\label{1.26}
{\zeta}_{\theta_n}=\frac{{\zeta}_{\theta_a}}{\beta}, \quad {\zeta}_{z_n}=\frac{{\zeta}_{z_a}}{\beta}
\end{equation}

The dimension of the constant coefficient $K_0$ in equation (\ref{1.14}) is calculated as \cite{na, na1, vis}
\begin{equation}\label{1.27}
[K_0]=M^{-\alpha-\delta-q} L^{3\alpha+\delta-s-1} T^{2\delta+3q-l}.
\end{equation}
 
 By following the approach of Sedov \cite{Sedov}, we get the conditions under which the formulated problem has self similar solutions. The relation between ${\rho}_a$, $p_a$ and $F_a$ is given as
\begin{equation}\label{1.28}
F_a= {p_a}^{3/2} {{\rho}_a}^{-1/2}.
\end{equation}

For the existence of similarity solution, the radiation absorption $K_0$ must depend on the dimensions of $F_a$ and ${\rho}_a$, which holds under the condition $s+l=-1$. 

\section{Similarity Transformations}
Zel'dovich and Raizer \cite{zel} showed that the gas dynamic equations acknowledge similarity transformations, that there are possible different flows similar to each other which are derivable from each other by changing the basic scales of length, time, and density. For self-similar motions, the system of fundamental partial differential equations (\ref{1.1})-(\ref{1.7}) reduces to a system of ordinary differential equations in new unknown functions of the similarity variable $\eta$ which is defined by 
\begin{equation*}
\eta=\frac{r}{R},\;\; R=R(t)=\bar{\beta} {F_a}^{1/3} {{\rho}_a}^{1/3} t.
\end{equation*}
The value of the constant $\bar{\beta}$ is so chosen that $\eta=1$ at the shock surface \cite{Sedov}.
 
The velocity, density, pressure, heat flux and length scales are not all independent of each other. If we choose $R$ and $\rho_{a}$ as the basic scales, then the quantity $\frac{dR}{dT}=\dot{R}$ can serve as the velocity scale, $\rho_{a} {\dot{R}}^2$ as the pressure scale. This does not restrict the universality of the solution as a scale is only defined within a numerical coefficient, which can always be involved in the new unknown function. Therefore, we represent the solution of the partial differential equations (\ref{1.1})-(\ref{1.7}) in terms of products of scale functions and the new unknown functions of the similarity variable $\eta$ in the following form \cite{vn, rangarao}
\begin{eqnarray}
&& u= \dot{R} U(\eta),\label{1.30}\\
&& v= \dot{R} V(\eta),\label{1.31}\\
&& w= \dot{R} W(\eta),\label{1.32}\\
&& \rho={\rho}_a D(\eta),\label{1.33}\\
&& p={\rho}_a {\dot{R}}^2 P(\eta),\label{1.34}\\
&& F=F_a \phi(\eta),\label{1.35}\\
&& Z=Z_a D(\eta),\label{1.36}\\
&& m=m_a S(\eta),\label{1.37}
\end{eqnarray}
where $U$, $V$, $W$, $D$, $P$, $\phi$ and $S$ are new non-dimensional functions of the similarity variable $\eta$. The differential equations are to be formulated in terms of the similarity variable $\eta$.
 
In order to obtain similarity solutions, the Shock Mach number $M$, which occurs in the shock conditions (\ref{1.241})-(\ref{1.248}) must be a constant parameter (i.e. Mach number should be independent of time). Using the equations (\ref{1.16}) and (\ref{1.20}) into $M=(\frac{{\dot{R}}^2 \rho_a}{\gamma p_a})^{1/2}$, we have obtained the expression for Shock Mach number as
\begin{equation}
M=\frac{2 {\dot{R}}^2 }{\gamma [\frac{{v^{*}}^2 R^{2\lambda}}{\lambda}-\bar{G} \pi {\rho}_a R^2]}.\label{1.38}
\end{equation}

Therefore, Mach number $M$ is constant for
\begin{equation}
\dot{R}=QR; \lambda=1,\label{1.39}
\end{equation}
and, we obtain a relation for gravitation parameter $G_0=\frac{\bar{G} \pi {\rho}_a}{Q^2}$ as
\begin{equation}
G_0=\frac{{v^{*}}^2}{Q^2}-\frac{2}{\gamma M^2}.\label{1.40}
\end{equation}

The relation (\ref{1.40}) is analogous to the relations (103) of Rogers \cite{rogers} and (20) of Singh \cite{ss} for the case of a perfect gas with a variable initial density of the medium, equation (3.7) of Patel \cite{patel} for the case of a mixture of perfect gas and small solid particles, equation (55) of Bajargaan and Patel \cite{ruchi} for the case of a  mixture of non-ideal gas and small solid particles. The quantity $G_0$ is a gravitation parameter, which is analogous to the parameter $l_1$ of Rosenau \cite{ros}. The parameter $\frac{v^*}{Q}$ is similar to the parameter $\frac{B}{Q}$ in relation (66) of Vishwakarma and Nath \cite{vv} in the absence of the gravitational field.

By using the similarity transformations (\ref{1.30})-(\ref{1.37}), the fundamental system of partial differential equations (\ref{1.1})-(\ref{1.7}) reduces into
\begin{eqnarray}
&& (U-\eta)\frac{dD}{d\eta}+D\frac{dU}{d\eta}+\frac{DU}{\eta}=0,\label{1.401}\\
&& U+(U-\eta)\frac{dU}{d\eta}+\frac{1}{D}\frac{dP}{d\eta}+G_0\frac{S}{\eta}-\frac{V^2}{\eta}=0,\label{1.402}\\
&& (U-\eta)\frac{dV}{d\eta}+V+\frac{UV}{\eta}=0,\label{1.403}\\
&& (U-\eta)\frac{dW}{d\eta}+W=0,\label{1.404}\\
&& \frac{dS}{d\eta}=2D\eta,\label{1.405}\\
&& 2P D(1-Z_a D)\{1+\bar{b}D(1-K_p)\}+(U-\eta) \bar{b} P D^2 D^{'}(1-K_p)\{Z_a+\bar{b}(1-K_p)\}\nonumber\\
&& {}-\Gamma P D^{'}(U-\eta)\{1+\bar{b}D(1-K_p)\}^2+P^{'} D (U-\eta)(1-Z_a D)\{1+\bar{b}D(1-K_p)\}\nonumber\\
&&{}=\frac{(\Gamma-1)D \{1+\bar{b}D(1-K_p)\}^2}{\eta \gamma^{3/2} M^3}(\eta \phi^{'}+\phi),\label{1.406}\\
&& \phi^{'}=(\gamma M^2)^{\delta} \xi {\eta}^s D^{\alpha} P^{\delta} {\phi}^{q+1},\label{1.407}
\end{eqnarray}
where
\begin{equation}
\xi=K_0 {\rho_a}^{\alpha-\frac{\delta}{3}-q} {\bar{\beta}}^{s+1},
\end{equation}
under the condition $3q+2\delta+s+1=0$ for similarity solutions. The quantity $\xi$ is a dimensionless constant taken as the parameter which characterizes the interaction between the gas and the incident radiation flux \cite{khud, na1}.

The above set of differential equations (\ref{1.401})-(\ref{1.407}) can be transformed and simplified into
\begin{eqnarray}
&& U^{'}=-\frac{U}{\eta}-\frac{(U-\eta)D^{'}}{D},\label{5.1}\\
&& D^{'}=\frac{1}{N}[D^2 (U-\eta)(1-{Z_a} D){\{1+\bar{b} D(1-K_p)\}}{\{U-\frac{(U-\eta)U}{\eta}+G_0 \frac{S}{\eta}-\frac{V^2 }{\eta}\}}-2 P D(1-{Z_a} D)\nonumber\\
&&{}\times{\{1+\bar{b} D(1-K_p)\}}+\frac{(\Gamma-1)D{\{1+\bar{b}D(1-K_P)\}}^2 \phi}{\eta \gamma^{3/2} M^3}{\{\eta (\gamma M^2)^{\delta} \xi {\eta}^{s} D^{\alpha} P^{\delta} \phi^{q}+1\}}],\label{5.2}\\
&& P^{'}=\frac{(U-\eta)U D}{\eta}-U D+(U-\eta)^2 D^{'}-\frac{G_0 S D}{\eta}+\frac{V^2 D}{\eta},\label{5.3}\\
&& V^{'}=-\frac{V}{(U-\eta)}-\frac{UV}{(U-\eta)\eta},\label{5.4}\\
&& W^{'}=-\frac{W}{(U-\eta)},\label{5.5}\\
&& S^{'}=2D\eta,\label{5.6}\\
&& \phi^{'}=(\gamma M^2)^{\delta} \xi {\eta}^s D^{\alpha} P^{\delta} {\phi}^{q+1},\label{5.7}
\end{eqnarray}
where
$N=(U-\eta)\bar{b}P D^2 (1-K_p)\{Z_a+\bar{b}(1-K_p)\}-(U-\eta)\Gamma P {\{1+\bar{b}D(1-K_p)\}}^2\\
{}+(U-\eta)^3 D(1-Z_a D)\{1+\bar{b}D(1-K_p)\}$,

By using the similarity transformations (\ref{1.30})-(\ref{1.37}), the shock conditions (\ref{1.241})-(\ref{1.248}) are transformed into
\begin{eqnarray}
&& U(1)=(1-\beta),\label{5.8}\\
&& D(1)=\frac{1}{\beta},\label{5.9}\\
&& P(1)=(1-\beta)+\frac{1}{\gamma M^2},\label{5.10}\\
&& V(1)=(G_0+\frac{2}{\gamma M^2})^{1/2},\label{5.11}\\
&& W(1)=\frac{w^{*}}{Q},\label{5.12}\\
&& S(1)=1,\label{5.13}\\
&& \phi(1)=1,\label{5.14}
\end{eqnarray}
where $\lambda=\sigma=1$.

In addition to shock conditions (\ref{5.8}) to (\ref{5.14}), the condition which is to be satisfied at the piston surface is that the velocity of the fluid is equal to the velocity of the piston itself. This kinematic condition from equation (\ref{1.30}) can be written as
\begin{equation}\label{5.1114}
U(\eta_p)=\eta_p,
\end{equation}
where $\eta_p=\frac{u_p}{\dot{R}}$.

After applying the similarity transformations (\ref{1.31}), (\ref{1.32}) on the equation (\ref{1.9}), the non-dimensional components of the vorticity vector $l_{r}=\frac{\zeta_{r}}{(\dot{R}/R)}$,$l_{\theta}=\frac{\zeta_{\theta}}{(\dot{R}/R)}$,$l_{z^{*}}=\frac{\zeta_{z^{*}}}{(\dot{R}/R)}$ in the flow-field behind the shock front can be written as
\begin{eqnarray}
&&l_{r}=0,\\
&&l_{\theta}=\frac{W}{2(U-\eta)},\\
&&l_{z^{*}}=-\frac{V}{(U-\eta)}.
\end{eqnarray}

For an isentropic change of the state of the mixture of the non-ideal gas and small solid particles, under the thermodynamic equilibrium condition, we may calculate the equilibrium sound speed in the mixture, as follows

\begin{equation}\label{4.340}
a_{m}=\biggl(\frac{\partial p}{\partial\rho}\biggr)_{S}^{\frac{1}{2}}=\biggl[\frac{\{\Gamma+(2\Gamma-Z)b\rho(1-K_{p})\}p}{(1-Z)\rho\{1+b\rho(1-K_{p})\}}\biggr]^{1/2},
\end{equation}

neglecting $b^{2}\rho^{2}$, where subscript `$S$' refers to the process of constant entropy. 

The adiabatic compressibility of the mixture of the non-ideal gas and small solid particles can be calculated as (c. f. \cite{moel})
\vspace{0.1mm}
\begin{eqnarray}
C_{adi}&&=-\rho\biggl(\frac{\partial}{\partial p}\biggl(\frac{1}{\rho}\biggr)\biggr)_{S}=\frac{1}{\rho {a_{m}}^{2}}\nonumber\\
&&{}=\frac{(1-Z)[1+b\rho(1-K_{p})]}{[\Gamma+(2\Gamma-Z)b\rho(1-K_{p})]p}.\label{4.343}
\end{eqnarray}

Using the equations (\ref{1.33}), (\ref{1.34}) and (\ref{1.36}) in the equation (\ref{4.343}), we get the non-dimensional expression for the adiabatic compressibility as

\begin{equation}\label{4.344}
(C_{adi})p_{a}=\frac{(1-Z_{a}D)[1+\bar{b}D(1-K_{p})]}{[\Gamma+\bar{b}D(1-K_{p})(2\Gamma-Z_{a}D)]\gamma M^{2}P}.
\end{equation}

Also, the total energy of the flow field between the piston and the cylindrical shock wave is given by
\begin{equation}\label{1}
E=2\pi \int\limits_{r_p}^{R} {\rho[\frac{1}{2}(u^{2}+v^{2}+w^{2})+U_{m}-\bar{G}m]r dr},
\end{equation}
where $r_p$ is the radius of the piston or the inner expanding surface . Now by using the similarity transformations (\ref{1.30}) to (\ref{1.37}) and the equations (\ref{1.39}) and (\ref{1.40}) in the relation (\ref{1}), we get

\begin{equation}\label{2}
E =2\pi {\rho}_{a} R^{4} Q^{2} J ,
\end{equation}
where
\begin{align}
J= &\int\limits_{\eta_{p}}^{1} {D[\frac{(U^{2}+V^{2}+W^{2})}{2}+\frac{P(1-Z_{a}D)}{(\Gamma-1)[1+\bar{b}D(1-K_{p})]}-G_{0} S]\eta d\eta},
\end{align}

$\eta_{p}$ being the value of `$\eta$' at the piston or inner expanding surface.

 Equation (\ref{2}) show that the total energy of the flow field behind the cylindrical shock wave is proportional to the fourth power of the shock radius $R$. This increase can be achieved by the pressure exerted on the fluid by the inner expanding surface (a contact surface or a piston). The situation of the similar type may prevail in the formation of cylindrical spark channels from exploding wires. In addition, in the usual cases of spark break down, time dependent energy input is a more natural assumption than the instantaneous energy input (c.f. \cite{vv, free}). It is also dependent on the gravitation parameter $G_{0}$ (c.f. \cite{ruchi}).

The ordinary differential equations (\ref{5.1})-(\ref{5.7}) with boundary conditions (\ref{5.8})-(\ref{5.14}) can now be numerically integrated to obtain the solution for the flow behind the shock surface.

Normalizing the variables $u$, $v$, $w$, $\rho$, $p$, $m$ and $F$ with their respective values at the shock, we obtain
\vspace{0.1mm}
\begin{eqnarray*}
 &&\frac{u}{u_{n}}=\frac{U(\eta)}{U(1)},\quad \frac{v}{v_{n}}=\frac{V(\eta)}{V(1)},\quad \frac{w}{w_{n}}=\frac{W(\eta)}{W(1)},\\
 &&\frac{\rho}{\rho_{n}}=\frac{D(\eta)}{D(1)}, \quad \frac{p}{p_{n}}=\frac{P(\eta)}{P(1)}, \quad \frac{m}{m_{n}}=\frac{\Omega(\eta)}{\Omega(1)},\\ 
 &&\frac{F}{F_{n}}=\frac{\phi(\eta)}{\phi(1)}.
\end{eqnarray*}

\section{Results and discussion}
For the existence of similarity solution of the present problem, the following conditions must be satisfied
\begin{equation*}
3q+2\delta+s+1=0,\; \text{and} \;\lambda=\sigma=1.
\end{equation*}
From the equation (\ref{1.22}), under the condition $\lambda=1$, the initial angular velocity $A_{a}$ become constant. Therefore, the similarity solution of present problem exist under variable azimuthal and axial fluid velocity and constant angular velocity. The total energy of the flow field behind the cylindrical shock wave is not constant but is proportional to the fourth power of the shock radius.

The distribution of the flow variables between the shock front $(\eta=1)$ and the inner expanding surface or piston $(\eta=\eta_{p})$ is obtained by numerical integration of the equations (\ref{5.1})-(\ref{5.7}) with the boundary conditions (\ref{5.8}) to (\ref{5.14}) by using the Runge-Kutta method of fourth order. The values of the constant parameters are taken to be $\gamma=1.4$; $K_{p}=0, 0.2, 0.4$; $G_{a}=50, 100$; $\beta^{'}=1$; $\bar{b}=0, 0.05, 0.1$; $M^{2}=25$; $G_{0}=0.25, 1, 5, 10, 20$ and $w^{*}/Q=0.005$, $\xi=0.1, 1, 10, 50$. The values $\gamma=1.4$; $\beta^{'}=1$ correspond to the mixture of air and glass particles \cite{miura}.  The value $M=5$ of the shock Mach number is appropriate, because we have treated the flow of a non-ideal gas and a pseudo-fluid (small solid particles) at a velocity and temperature equilibrium. The value $K_p=0$ corresponds to the dust-free case and $K_p=0$, $\bar{b}=0$ correspond to the perfect gas case.

The values of the density ratio $\beta$ across the shock front, shock strength $(1-\beta)$ and position of inner expanding surface or piston $\eta_{p}$ are tabulated in table $1$, $2$ and $3$ for different values of $K_p$, $G_a$, $\bar{b}$; $K_p$, $G_a$, $\bar{b}$, $\xi$, and $G_0$ respectively. Figures $1$, $2$ and $3$ show the variation of the reduced flow variables $u/u_n$, $v/v_{n}$, $w/w_{n}$, $\rho/\rho_{n}$, $p/p_{n}$, $m/m_{n}$, $F/F_{n}$, $l_{\theta}/l_{\theta_n}$, $l_{z^{*}}/l_{z^{*}_{n}}$ and the reduced adiabatic compressibility $C_{adi}/{(C_{adi})}_n$ with $\eta$ for various values of parameters $K_p$, $\bar{b}$, $G_a$; $K_p$, $G_a$, $\bar{b}$, $\xi$, and $G_0$ respectively and fixed values of other constant parameters. It is evident from the figures $1$, $2$ and $3$ that as we move from the inner expanding surface (piston) towards the shock front, the radial component of fluid velocity $u/u_{n}$, the density $\rho/\rho_{n}$, the pressure $p/p_{n}$, the azimuthal component of vorticity vector $l_{\theta}/l_{\theta_n}$, the axial component of vorticity vector $l_{z^{*}}/l_{z^{*}_{n}}$ decrease and azimuthal component of fluid velocity $v/v_{n}$, the axial component of fluid velocity $w/w_{n}$, the mass $m/m_{n}$, the total heat flux $F/F_n$ and the adiabatic compressibility $C_{adi}/{(C_{adi})}_n$ increase.  
\vskip1.2em
The effects of an increase in the value of the non-idealness parameter $\bar{b}$ are manifested as follows:
\begin{description}
\item[(i)] the value of the shock strength decreases (see Table $1$),
\item[(ii)] the distance $(1-{\eta}_{p})$ between the piston and the shock front increases (see Table $1$),
\item[(iii)] the flow variables $u/u_{n}$, $\rho/\rho_{n}$, $p/p_{n}$, $F/F_{n}$, $l_{\theta}/l_{\theta_{n}}$, $l_{z^{*}}/l_{z^{*}_{n}}$ decrease (see figure $1$),
\item[(iv)] the flow variables $v/v_{n}$, $w/w_{n}$, $m/m_{n}$, $C_{adi}/(C_{adi})_n$ increase (see figure $1$).
\end{description}

These effects may be physically interpreted as follows:

An increase in non-idealness parameter $\bar{b}$ increased the compressibility of the medium. This gives an increase in the distance between the shock front and the piston, a decrease in the shock strength, and the above behaviour of the flow variables.

\vskip1.2em
The effects of an increase in the value of mass concentration of solid particles $K_{p}$ in the mixture are given as follows:
\begin{description}
\item[(i)] the value of the shock strength increases (see Table $1$),
\item[(ii)] the distance $(1-{\eta}_{p})$ between the piston and the shock front decreases (see Table $1$),
\item[(iii)] the flow variables $u/u_{n}$, $v/v_{n}$, $w/w_{n}$, $m/m_{n}$, $(C_{adi})/(C_{adi})_n$ decrease (see figure $1$),
\item[(iv)] the flow variables $\rho/\rho_{n}$, $p/p_{n}$, $F/F_{n}$, $l_{\theta}/l_{\theta_{n}}$, $l_{z^{*}}/l_{z^{*}_{n}}$ increase (see figure $1$).
\end{description}

The physical interpretation of these effects can be formulated as follows: Small solid particles of density equal or greater than of the perfect gas (non-ideal gas) in the mixture occupy a significant portion of the volume, which lowers the compressibility of the medium. Then, an increase in $K_p$ further reduces the compressibility, which causes a decrease in the shock strength and the above behaviour of the flow variables.

The effects of an increase in the ratio of the density of the solid particles to the initial density of the gas $G_{a}$ are listed as follows:
\begin{description}
\item[(i)] the value of the shock strength increases (see Table $1$),
\item[(ii)] the distance $(1-{\eta}_{p})$ between the piston and the shock front decreases (see Table $1$),
\item[(iii)] the flow variables $u/u_{n}$, $v/v_{n}$, $(C_{adi})/(C_{adi})_n$ have negligible effects (see figure $1$),
\item[(iv)] for the case of $K_{p}=0.2$, the flow variables $w/w_{n}$, $\rho/\rho_{n}$, $p/p_{n}$, $m/m_{n}$, $F/F_{n}$, $l_{\theta}/l_{\theta_{n}}$, $l_{z^{*}}/l_{z^{*}_{n}}$, $a_{iso}/(a_{iso})_n$ have negligible effects but for the case of $K_{p}=0.4$, the flow variables $w/w_{n}$, $m/m_{n}$, $a_{iso}/(a_{iso})_n$ decrease and the flow variables $\rho/\rho_{n}$, $p/p_{n}$, $F/F_{n}$, $l_{\theta}/l_{\theta_{n}}$, $l_{z^{*}}/l_{z^{*}_{n}}$ increase (see figure $1$).  
\end{description}

The above effects are more impressive at higher values of $K_p$. These effects may be physically interpreted as follows:

By an increase in $G_a$ (at constant $K_p$), there is high decrease in $Z_a$, i.e., the volume fraction of solid particles in the undisturbed medium becomes, comparatively, very small. This causes comparatively more compression of the mixture in the region between the shock and inner expanding surface, which displays the above effects.

The effects of an increase in the radiation parameter $\xi$ are 
\begin{description}
\item[(i)] to increase the distance $(1-{\eta}_{p})$ between the piston and the shock front (see Table $2$),
\item[(ii)] to have negligible effects on the flow variables $u/u_{n}$, $v/v_{n}$, $w/w_{n}$, $\rho/\rho_{n}$, $p/p_{n}$, $m/m_{n}$, $l_{\theta}/l_{\theta_{n}}$, $l_{z^{*}}/l_{z^{*}_{n}}$, $(C_{adi})/(C_{adi})_n$ (see figure $2$),
\item[(iii)] to decrease the radiation heat flux $F/F_{n}$. The radiation parameter $\xi$ dominates the effect of all the dusty gas parameters such as $\bar{b}$, $K_p$ and $G_a$ for $\xi$ from $1$ to $50$. For $\xi=10,$ and $50$, the radiation flux increases rapidly near the shock front (see figure $2(g)$),
\end{description}
and the shock strength remains constant.

The above effects show that the monochromatic radiation is more absorbed by the gas in the flow-field behind the shock front by an increase in the radiation parameter $\xi$.

The effects of an increase in the value of the gravitational parameter $G_{0}$ on flow variables are listed as follows:
\begin{description}
\item[(i)] the distance $(1-{\eta}_{p})$ between the piston and the shock front decreases (see Table $3$),
\item[(ii)] the flow variables $u/u_{n}$, $\rho/\rho_{n}$, $p/p_{n}$, $F/F_{n}$, $l_{\theta}/l_{\theta_{n}}$, $l_{z^{*}}/l_{z^{*}_{n}}$ increase (see Figure $3$),
\item[(iii)] the flow variables $v/v_{n}$, $w/w_{n}$, $m/m_{n}$, $(C_{adi})/(C_{adi})_n$ decrease (see Figure $3$),
\end{description}
and the shock strength remains constant.  

As the gravitation parameter $G_0=\frac{\bar{G} \pi {\rho}_{a}}{Q^2}$ is inversely proportional to $Q$ and $Q$ is proportional to the shock velocity $\dot{R}$, therefore by increasing the value of gravitation parameter $G_0$, the shock velocity $\dot{R}$ decreases. Now, from equation (\ref{5.1114}), position of the piston ${\eta}_{p}$ is inversely proportional to $\dot{R}$, therefore ${\eta}_{p}$ increases with $G_0$. So, the distance between the piston and the shock front ($1-{\eta}_{p}$) decreases. The compressibility of the medium decreases with $G_0$, therefore, we get the above variation of the flow variables.

\begin{center}
\bf{Table $1$}
\end{center}
\textbf{Variation of the density ratio $\beta(=\frac{\rho_{a}}{\rho_{n}})$ across the shock front and the position of the piston $\eta_{p}$ for different values of $K_{p}$, $G_{a}$ and $\bar{b}$ with $\beta^{'}=1$, $\gamma=1.4$, $M=5$, $\frac{w^{*}}{Q}=0.005$, $\xi=0.1$, $\alpha=-0.5$, $\delta=-1$, $q=0$, $s=1$, and $G_{0}=1$.}

\begin{center}\small
\begin{tabular}{|l|c|c|c|c|c|c|r|}
\hline
$K_{p}$ & $\Gamma$ & $G_{a}$ & $z_{a}$ & $\bar{b}$ & $\beta$ & $1-\beta=\frac{u_{n}}{\dot{R}}$ & position of the piston
$\eta_{p}=\frac{u_{p}}{\dot{R}}$ \\
\hline
0 & 1.4 & - & 0 & 0 & 0.2 & 0.8 & 0.951548 \\
  &  &  &  & 0.05 & 0.234135 & 0.765865 & 0.934964 \\
    &  &  &  & 0.1 & 0.259555 & 0.740445 & 0.923665 \\
  0.2 & 1.32  & 50 & 0.00497512 & 0 & 0.174732 & 0.825268 & 0.957461 \\
    &  &   &  & 0.05 &  0.202648 & 0.797352 & 0.944545 \\
    &  &   &  & 0.1 &  0.223904 & 0.776096 & 0.935478 \\
   &  & 100 & 0.00249377 & 0 & 0.172593 & 0.827407 & 0.958653 \\
     & & & & 0.05 & 0.200824 & 0.799176 & 0.945484\\
     & & & & 0.1 & 0.222237 & 0.777763 & 0.936343\\
    0.4 & 1.24 & 50 & 0.0131579 & 0 & 0.150524 & 0.849476 & 0.961961\\
     & & & & 0.05 & 0.171411 & 0.828589 & 0.952987\\
     & & & & 0.1 & 0.187857 & 0.812143 & 0.946322\\
     & & 100 & 0.00662252 & 0 & 0.144688 & 0.855312 & 0.962901\\
     & & & & 0.05 & 0.166339 & 0.833661 & 0.955595\\
     & & & & 0.1 & 0.183188 & 0.816812 & 0.948723\\

  \hline
\end{tabular}
\end{center}

\begin{center}
\bf{Table $2$}
\end{center}
\textbf{Variation of the density ratio $\beta(=\frac{\rho_{a}}{\rho_{n}})$ across the shock front and the position of the piston $\eta_{p}$ for different values of $K_{p}$, $G_{a}$, $\bar{b}$ and $\xi$ with $\beta^{'}=1$, $\gamma=1.4$, $M=5$, $\frac{w^{*}}{Q}=0.005$, $\alpha=-0.5$, $\delta=-1$, $q=0$, $s=1$, and $G_{0}=0.25$.}

\begin{center}\small
\begin{tabular}{|l|c|c|c|c|c|c|c|r|}
\hline
$K_{p}$ & $\Gamma$ & $G_{a}$ & $z_{a}$ & $\bar{b}$ & $\xi$ & $\beta$ & $1-\beta=\frac{u_{n}}{\dot{R}}$ & position of the piston
$\eta_{p}=\frac{u_{p}}{\dot{R}}$ \\
\hline
0 & 1.4 & - & 0 & 0 & 0.1 & 0.2 & 0.8 & 0.95096 \\
  & & & & & 1 & & & 0.95096\\
  & & & & & 10 & & & 0.95095\\
  & & & & & 50 & & & 0.950951\\
  & & & & 0.1 & 0.1 & 0.259555 & 0.740445 & 0.922959\\
  & & & & & 1 & & & 0.922959\\ 
  & & & & & 10 & & & 0.922955\\
  & & & & & 50 & & & 0.922937\\
0.2 & 1.32 & 50 & 0.00497512 & 0 & 0.1 & 0.174732 & 0.825268 & 0.956963\\
  & & & & & 1 & & & 0.956963\\
  & & & & & 10 & & & 0.956962\\
  & & & & & 50 & & & 0.956957\\
  & & 100 & 0.00249377 & & 0.1 & 0.172593 & 0.827407 & 0.958155\\
  & & & & & 1 & & & 0.958155\\
  & & & & & 10 & & & 0.958154\\
  & & & & & 50 & & & 0.958149\\
  & & 50 & 0.00497512 & 0.1 & 0.1 & 0.223904 & 0.776096 & 0.934875\\
  & & & & & 1 & & & 0.934875\\
  & & & & & 10 & & & 0.934873\\
  & & & & & 50 & & & 0.934862\\
  & & 100 & 0.00249377 & & 0.1 & 0.222237 & 0.777763 & 0.935742\\
  & & & & & 1 & & & 0.935742\\
  & & & & & 10 & & & 0.935739\\
  & & & & & 50 & & & 0.935729\\

  \hline
\end{tabular}
\end{center}

\begin{center}
\bf{Table $3$}
\end{center}
\textbf{Variation of the density ratio $\beta(=\frac{\rho_{a}}{\rho_{n}})$ across the shock front and the position of the piston $\eta_{p}$ for different values of $G_0$ with $\beta^{'}=1$, $\gamma=1.4$, $M=5$, $\frac{w^{*}}{Q}=0.005$, $\xi=10$, $\alpha=-0.5$, $\delta=-1$, $q=0$, $s=1$, $K_{p}=0.2$, $G_{a}=50$ and  $\bar{b}=0.01$.}

\begin{center}\small
\begin{tabular}{|l|c|c|r|}
\hline
  $G_0$ & $\beta$ & $1-\beta=\frac{u_{n}}{\dot{R}}$ & position of the piston
$\eta_{p}=\frac{u_{p}}{\dot{R}}$ \\
\hline
0.25 & 0.181142 & 0.818858 & 0.953751\\
1 & 0.181142 & 0.818858 & 0.954258\\
5 & 0.181142 & 0.818858 & 0.956601\\
10 & 0.181142 & 0.818858 & 0.958914\\
20 & 0.181142 & 0.818858 & 0.962321\\

  \hline
\end{tabular}
\end{center}

\begin{figure}
\begin{center}
\includegraphics[scale=0.6]{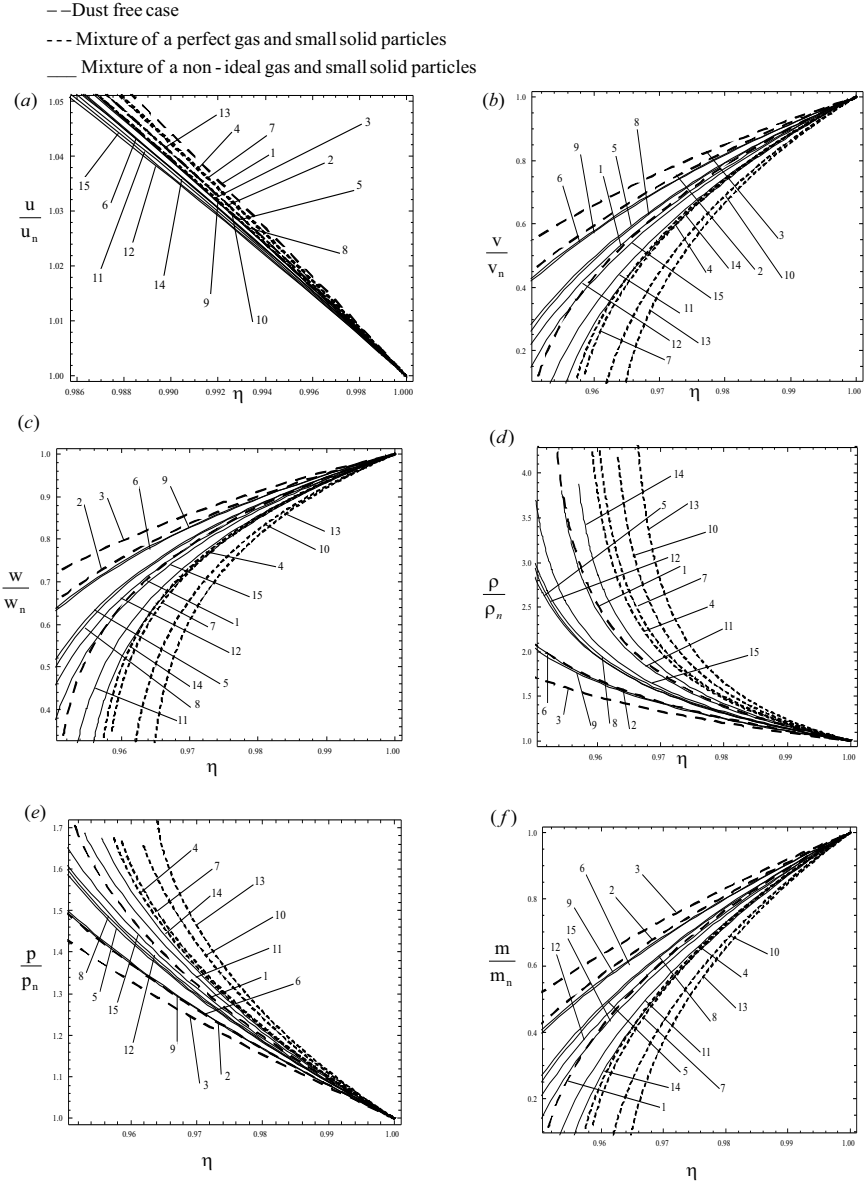}
\caption{{\bf{Fig. 1}} Variation of the flow variables (a) radial component of fluid velocity (b) azimuthal component of fluid velocity (c) axial component of fluid velocity (d) density (e) pressure (f) mass, in the region behind the shock front in case of $\xi=0.1$, $\frac{w^{*}}{Q}=0.005$, $\alpha=-0.5$, $\delta=-1$, $q=0$, $s=1$, $G_{0}=1$; 1. $K_{p}=0$, $\bar{b}=0$(perfect gas); 2. $K_{p}=0$, $\bar{b}=0.05$(non-ideal gas); 3. $K_{p}=0$, $\bar{b}=0.1$(non-ideal gas); 4. $K_{p}=0.2$, $\bar{b}=0$, $G_{a}=50$; 5.$K_{p}=0.2$, $\bar{b}=0.05$, $G_{a}=50$; 6. $K_{p}=0.2$, $\bar{b}=0.1$, $G_{a}=50$; 7. $K_{p}=0.2$, $\bar{b}=0$, $G_{a}=100$; 8. $K_{p}=0.2$, $\bar{b}=0.05$, $G_{a}=100$; 9. $K_{p}=0.2$, $\bar{b}=0.1$, $G_{a}=100$; 10. $K_{p}=0.4$, $\bar{b}=0$, $G_{a}=50$; 11. $K_{p}=0.4$, $\bar{b}=0.05$, $G_{a}=50$; 12. $K_{p}=0.4$, $\bar{b}=0.1$, $G_{a}=50$; 13. $K_{p}=0.4$, $\bar{b}=0$, $G_{a}=100$; 14. $K_{p}=0.4$, $\bar{b}=0.05$, $G_{a}=100$; 15. $K_{p}=0.4$, $\bar{b}=0.1$, $G_{a}=100$}

\end{center}
\end{figure}

\begin{figure}
\begin{center}
\includegraphics[scale=0.6]{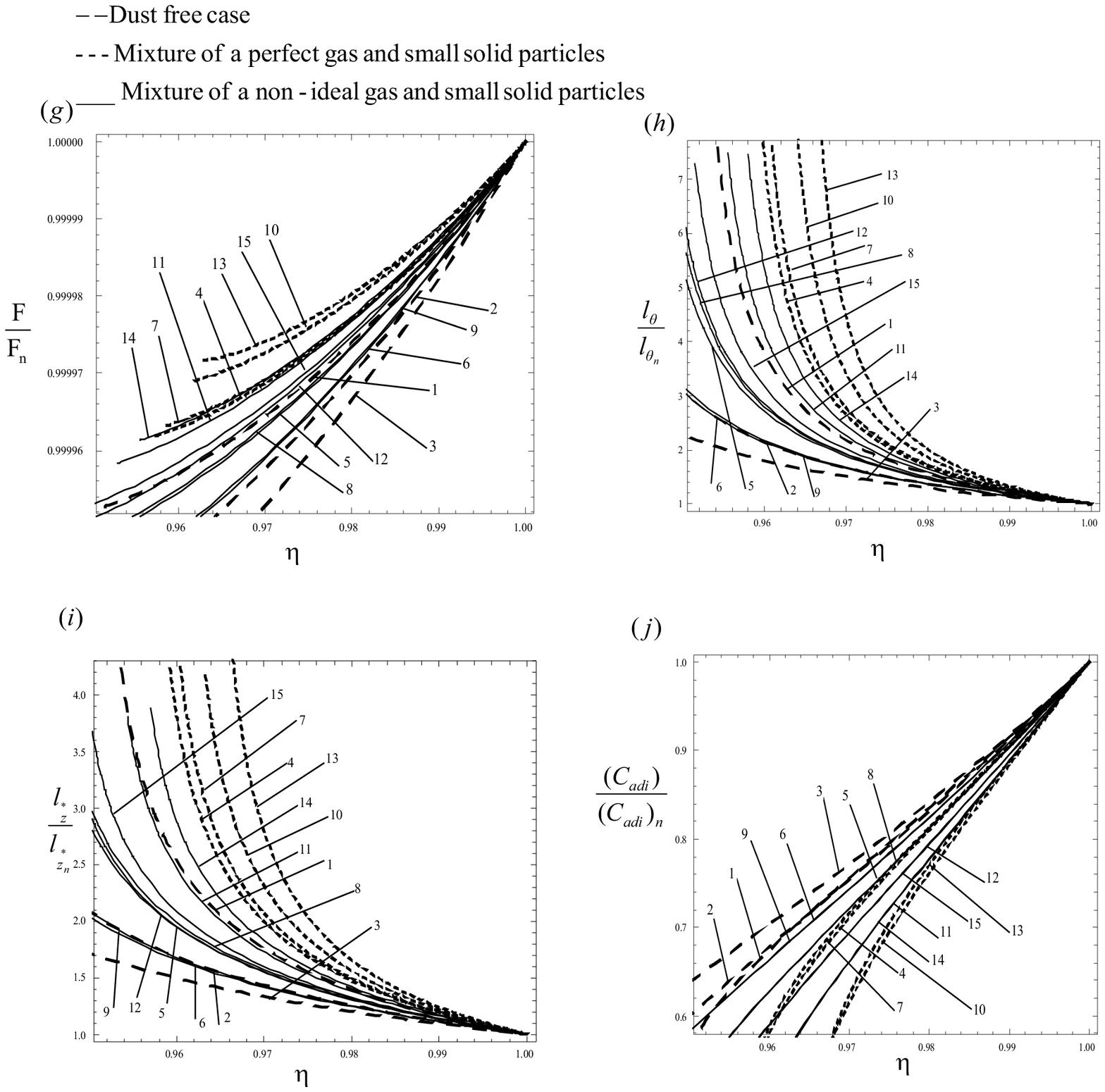}
\caption{{\bf{Fig. 1}} Variation of the flow variables (g) radiation flux (h) azimuthal component of vorticity vector (i) axial component of vorticity vector (j) adiabatic compressibility (k) isothermal speed of sound, in the region behind the shock front in case of $\xi=0.1$, $\frac{w^{*}}{Q}=0.005$, $\alpha=-0.5$, $\delta=-1$, $q=0$, $s=1$, $G_{0}=1$; 1. $K_{p}=0$, $\bar{b}=0$(perfect gas); 2. $K_{p}=0$, $\bar{b}=0.05$(non-ideal gas); 3. $K_{p}=0$, $\bar{b}=0.1$(non-ideal gas); 4. $K_{p}=0.2$, $\bar{b}=0$, $G_{a}=50$; 5.$K_{p}=0.2$, $\bar{b}=0.05$, $G_{a}=50$; 6. $K_{p}=0.2$, $\bar{b}=0.1$, $G_{a}=50$; 7. $K_{p}=0.2$, $\bar{b}=0$, $G_{a}=100$; 8. $K_{p}=0.2$, $\bar{b}=0.05$, $G_{a}=100$; 9. $K_{p}=0.2$, $\bar{b}=0.1$, $G_{a}=100$; 10. $K_{p}=0.4$, $\bar{b}=0$, $G_{a}=50$; 11. $K_{p}=0.4$, $\bar{b}=0.05$, $G_{a}=50$; 12. $K_{p}=0.4$, $\bar{b}=0.1$, $G_{a}=50$; 13. $K_{p}=0.4$, $\bar{b}=0$, $G_{a}=100$; 14. $K_{p}=0.4$, $\bar{b}=0.05$, $G_{a}=100$; 15. $K_{p}=0.4$, $\bar{b}=0.1$, $G_{a}=100$}
\end{center}
\end{figure}

\begin{figure}
\begin{center}
\includegraphics[scale=0.6]{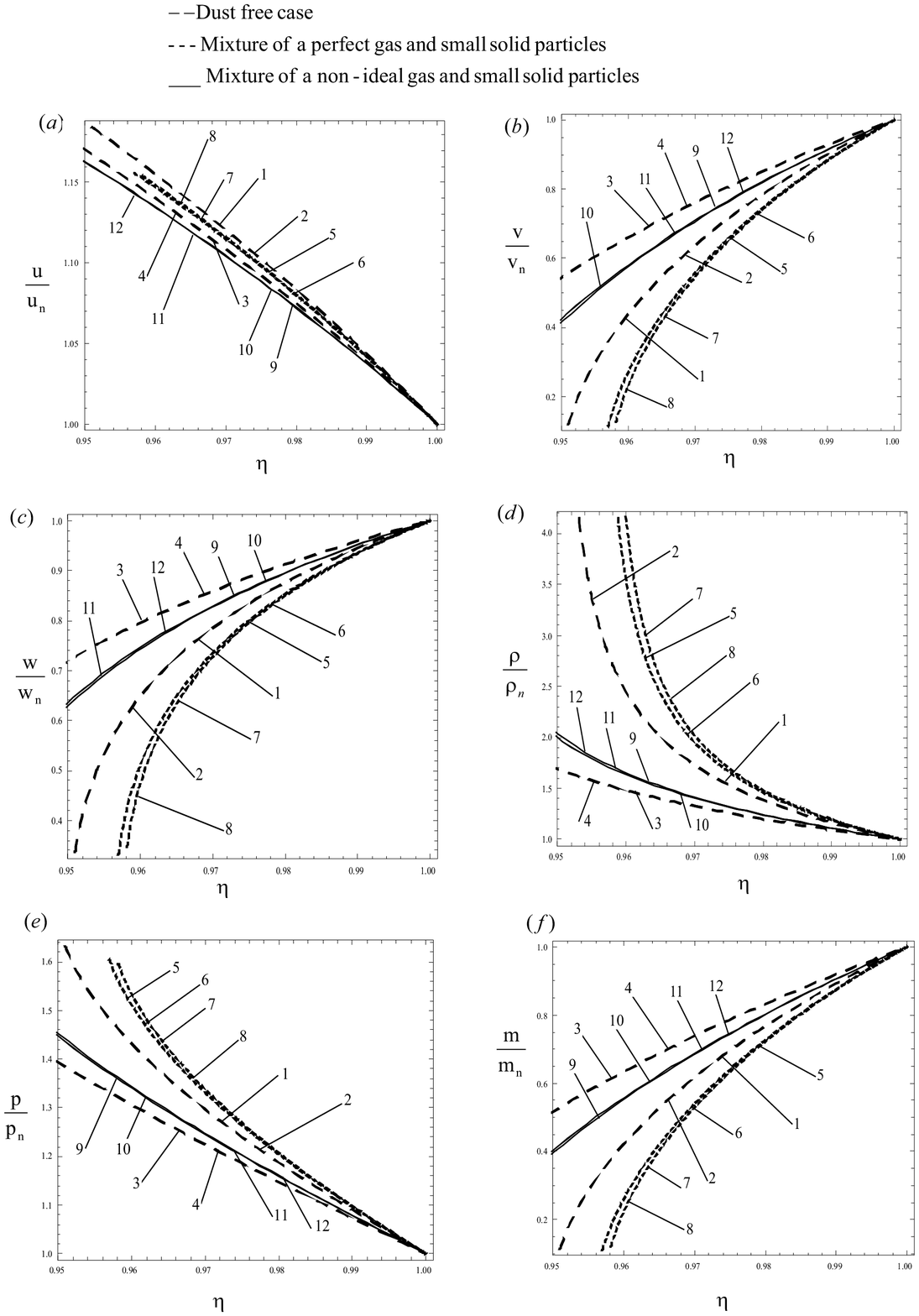}
\caption{{\bf{Fig. 2}} Variation of the flow variables (a) radial component of fluid velocity (b) azimuthal component of fluid velocity (c) axial component of fluid velocity (d) density (e) pressure (f) mass, in the region behind the shock front in case of $\frac{w^{*}}{Q}=0.005$, $\alpha=-0.5$, $\delta=-1$, $q=0$, $s=1$, $G_{0}=0.25$; 1. $K_{p}=0$, $\bar{b}=0$, $\xi=1$(perfect gas); 2. $K_{p}=0$, $\bar{b}=0$, $\xi=50$(perfect gas); 3. $K_{p}=0$, $\bar{b}=0.1$, $\xi=1$(non-ideal gas); 4. $K_{p}=0$, $\bar{b}=0.1$, $\xi=50$(non-ideal gas); 5.$K_{p}=0.2$, $\bar{b}=0$, $G_{a}=50$, $\xi=1$; 6. $K_{p}=0.2$, $\bar{b}=0$, $G_{a}=50$, $\xi=50$; 7. $K_{p}=0.2$, $\bar{b}=0$, $G_{a}=100$, $\xi=1$; 8. $K_{p}=0.2$, $\bar{b}=0$, $G_{a}=100$, $\xi=50$; 9. $K_{p}=0.2$, $\bar{b}=0.1$, $G_{a}=50$, $\xi=1$; 10. $K_{p}=0.2$, $\bar{b}=0.1$, $G_{a}=50$, $\xi=50$; 11. $K_{p}=0.2$, $\bar{b}=0.1$, $G_{a}=100$, $\xi=1$; 12. $K_{p}=0.2$, $\bar{b}=0.1$, $G_{a}=100$, $\xi=50$}

\end{center}
\end{figure}

\begin{figure}
\begin{center}
\includegraphics[scale=0.8]{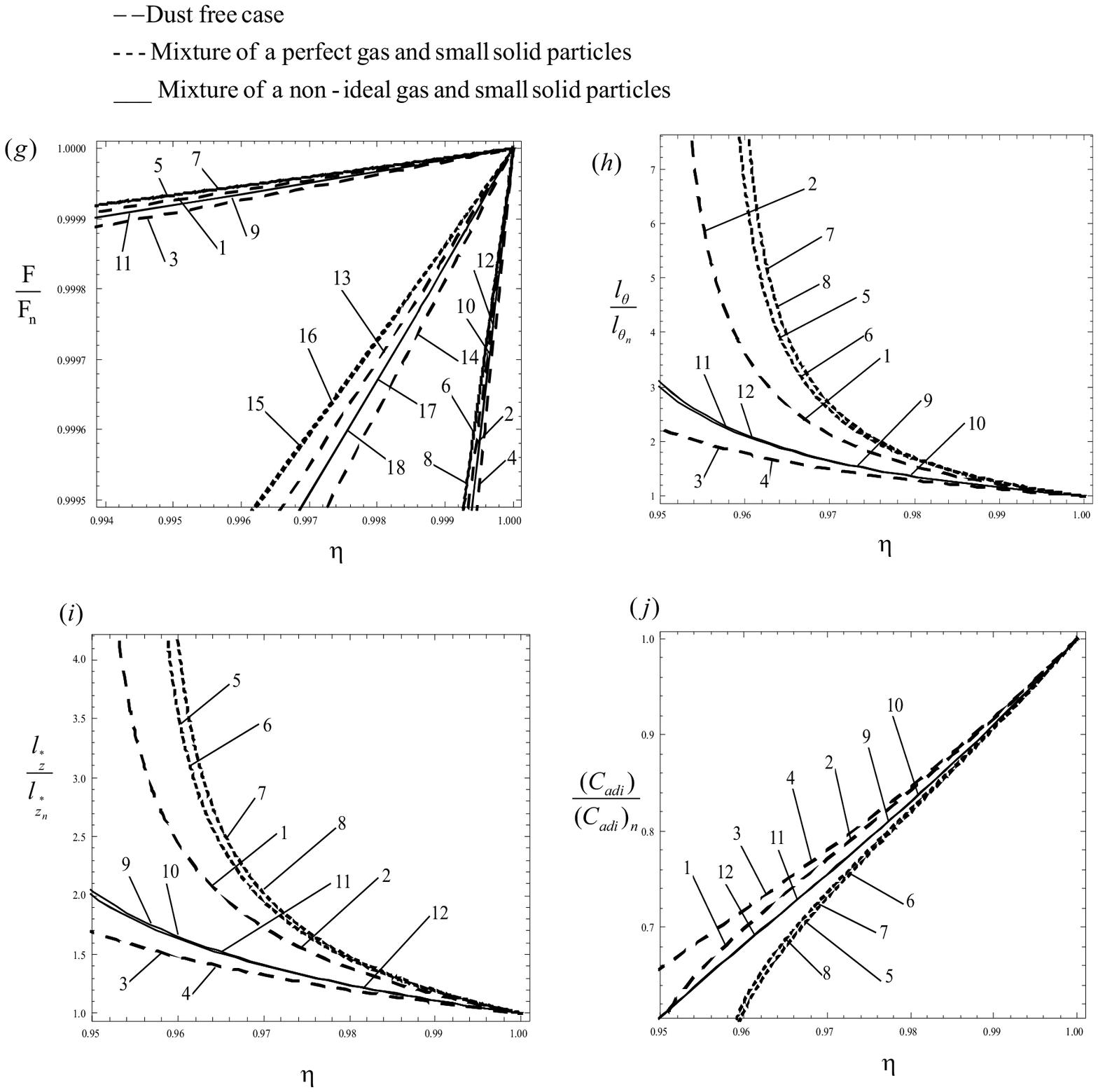}
\caption{{\bf{Fig. 2}} Variation of the flow variables (g) radiation flux (h) azimuthal component of vorticity vector (i) axial component of vorticity vector (j) adiabatic compressibility (k) isothermal speed of sound, in the region behind the shock front in case of $\frac{w^{*}}{Q}=0.005$, $\alpha=-0.5$, $\delta=-1$, $q=0$, $s=1$, $G_{0}=0.25$; 1. $K_{p}=0$, $\bar{b}=0$, $\xi=1$(perfect gas); 2. $K_{p}=0$, $\bar{b}=0$, $\xi=50$(perfect gas); 3. $K_{p}=0$, $\bar{b}=0.1$, $\xi=1$(non-ideal gas); 4. $K_{p}=0$, $\bar{b}=0.1$, $\xi=50$(non-ideal gas); 5.$K_{p}=0.2$, $\bar{b}=0$, $G_{a}=50$, $\xi=1$; 6. $K_{p}=0.2$, $\bar{b}=0$, $G_{a}=50$, $\xi=50$; 7. $K_{p}=0.2$, $\bar{b}=0$, $G_{a}=100$, $\xi=1$; 8. $K_{p}=0.2$, $\bar{b}=0$, $G_{a}=100$, $\xi=50$; 9. $K_{p}=0.2$, $\bar{b}=0.1$, $G_{a}=50$, $\xi=1$; 10. $K_{p}=0.2$, $\bar{b}=0.1$, $G_{a}=50$, $\xi=50$; 11. $K_{p}=0.2$, $\bar{b}=0.1$, $G_{a}=100$, $\xi=1$; 12. $K_{p}=0.2$, $\bar{b}=0.1$, $G_{a}=100$, $\xi=50$; 13. $K_{p}=0$, $\bar{b}=0$, $\xi=10$(perfect gas); 14. $K_{p}=0$, $\bar{b}=0.1$, $\xi=10$(non-ideal gas); 15. $K_{p}=0.2$, $\bar{b}=0$, $G_{a}=50$, $\xi=10$; 16. $K_{p}=0.2$, $\bar{b}=0$, $G_{a}=100$, $\xi=10$; 17. $K_{p}=0.2$, $\bar{b}=0.1$, $G_{a}=50$, $\xi=10$; 18. $K_{p}=0.2$, $\bar{b}=0.1$, $G_{a}=100$, $\xi=10$}

\end{center}
\end{figure}

\begin{figure}
\begin{center}
\includegraphics[scale=0.6]{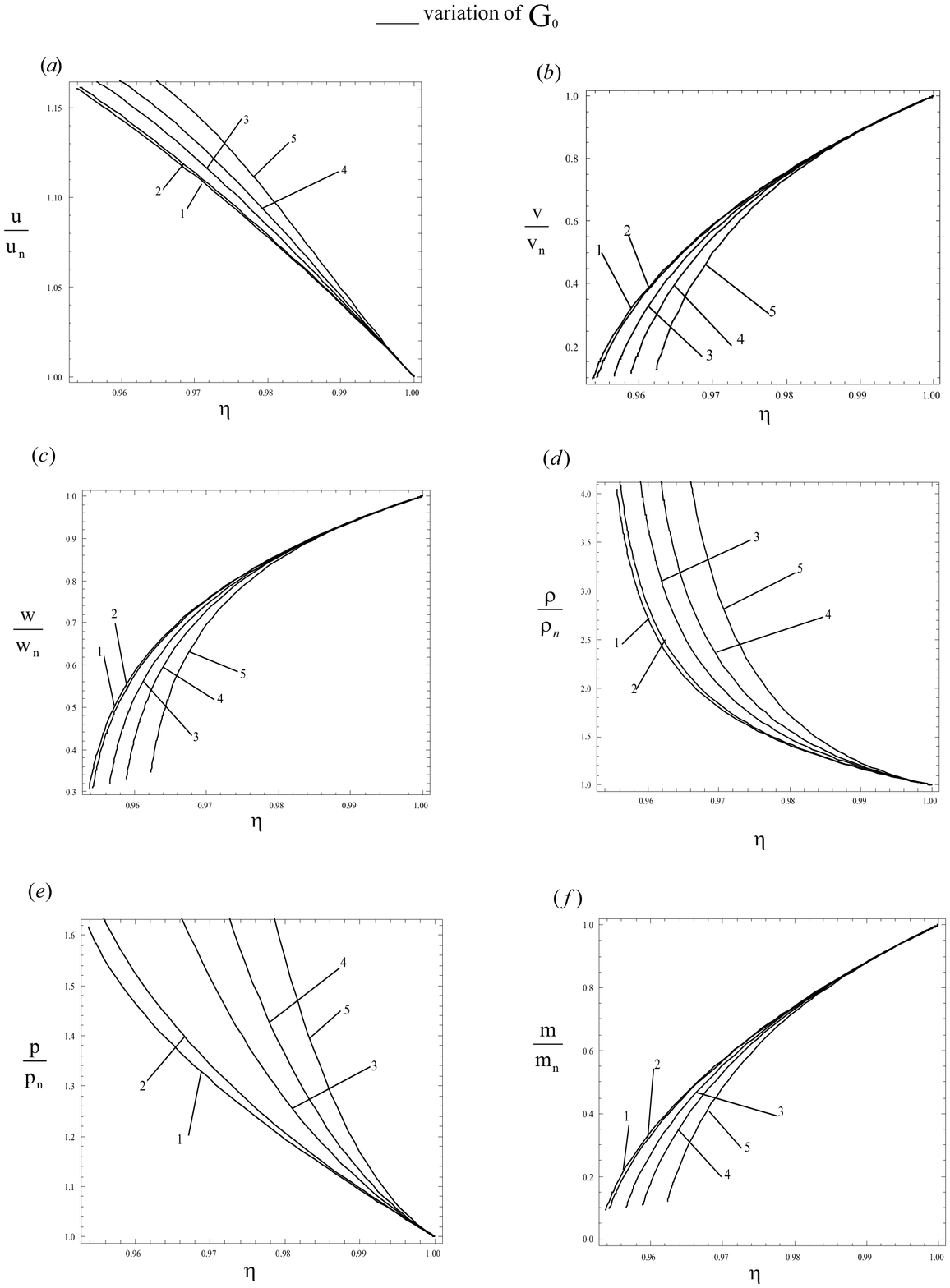}
\caption{{\bf{Fig. 3}} Variation of the flow variables (a) radial component of fluid velocity (b) azimuthal component of fluid velocity (c) axial component of fluid velocity (d) density (e) pressure (f) mass, in the region behind the shock front in case of $\frac{w^{*}}{Q}=0.005$, $\xi=10$, $K_{p}=0.2$, $\bar{b}=0.01$, $G_{a}=50$, $\alpha=-0.5$, $\delta=-1$, $q=0$, $s=1$; 1. $G_{0}=0.25$; 2. $G_{0}=1$; 3. $G_{0}=5$; 4. $G_{0}=10$; 5. $G_{0}=20$}
\end{center}
\end{figure}

\begin{figure}
\begin{center}
\includegraphics[scale=0.8]{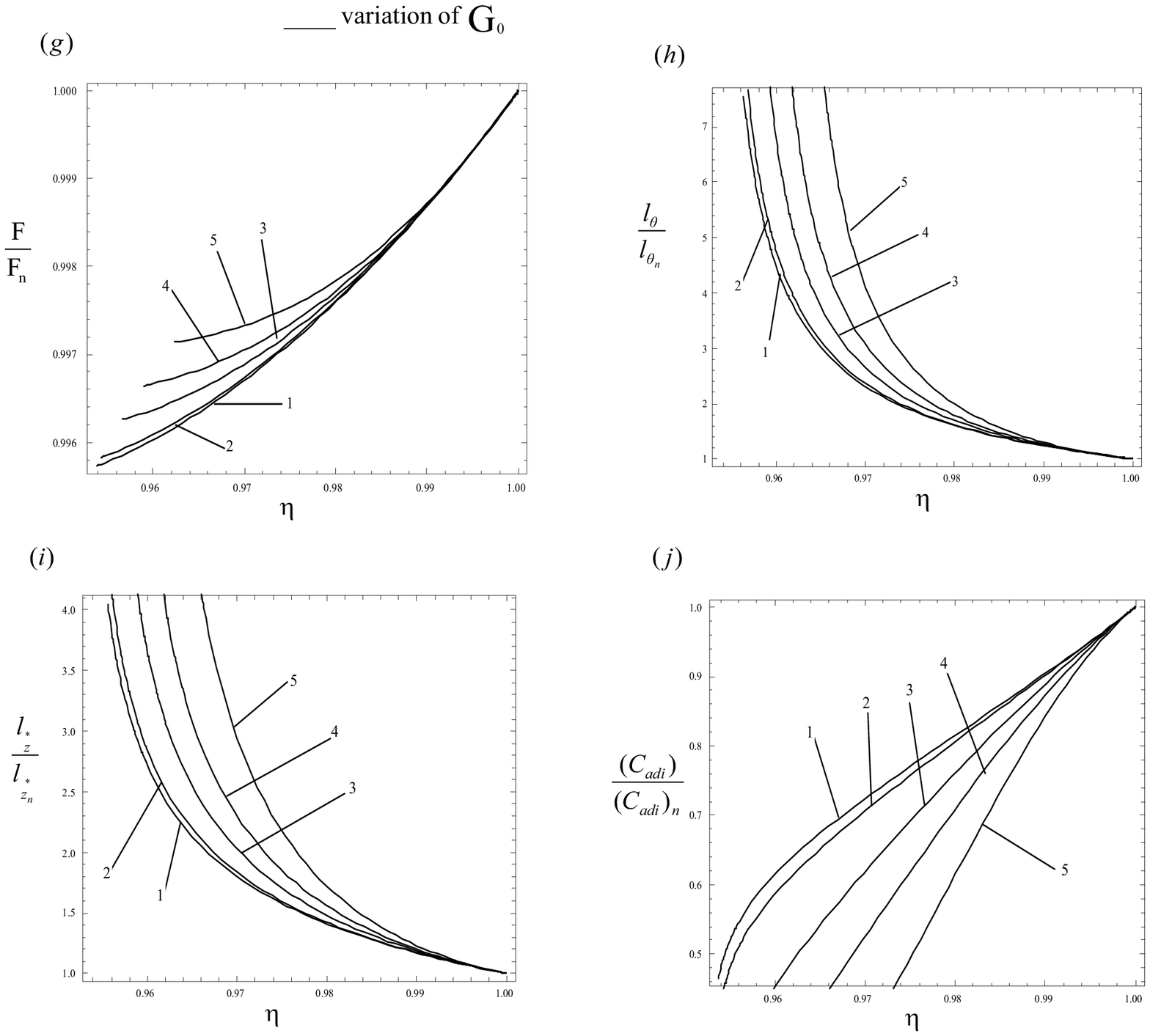}
\caption{{\bf{Fig. 3}} Variation of the flow variables (g) radiation flux (h) azimuthal component of vorticity vector (i) axial component of vorticity vector (j) adiabatic compressibility (k) isothermal speed of sound, in the region behind the shock front in case of $\frac{w^{*}}{Q}=0.005$, $\xi=10$, $K_{p}=0.2$, $\bar{b}=0.01$, $G_{a}=50$, $\alpha=-0.5$, $\delta=-1$, $q=0$, $s=1$; 1. $G_{0}=0.25$; 2. $G_{0}=1$; 3. $G_{0}=5$; 4. $G_{0}=10$; 5. $G_{0}=20$}
\end{center}
\end{figure}

\section{Conclusions}
The present work investigates the self similar flow behind a cylindrical shock wave propagating in a rotating axisymmetric mixture of non-ideal gas and small solid particles under the action of monochromatic radiation and gravitational field. The fluid velocities are assumed to vary and the initial density is assumed to be constant. The findings of this work provides a clear picture of whether and how the non-idealness parameter, the mass concentration of solid particles in the mixture, the ratio of the density of solid particles to the initial density of the gas, the radiation parameter and the gravitation parameter affect the shock strength and the fluid flow behind the shock front. On the basis of this work, one may draw the following important conclusions:
\begin{description}
\item[(i)] The similarity solution of the present problem exists only when the radiation exponent $q$ is dependent on the pressure exponent $\delta$ and position exponent $s$ or time exponent $l$ in the radiation absorption coefficient i.e. $3q+2\delta+s+1=0\;\&\;s+l=-1$, and the azimuthal fluid velocity exponent $\lambda$ is equal to the axial fluid velocity exponent $\sigma$ and their value is $1$. Under this condition, the initial angular velocity become constant.
\item[(ii)] The effects of the radiation parameter $\xi$ are negligible on the variation of the flow variables except the radiation heat flux, and it dominates the effect of dusty gas parameters on the variation of radiation heat flux behind the shock front.
\item[(iii)] The total energy of the flow-field behind the shock front is not constant but it is  proportional to the fourth power of the shock radius $R$.
\item[(iv)] The shock strength is independent from the radiation parameter $\xi$ and the gravitation parameter $G_{0}$ but it depends on the dusty gas parameters $\bar{b}$, $K_{p}$ and $G_{a}$.
\item[(v)] The distance between the piston and the shock front increases by increasing the value of the radiation parameter $\xi$, the non-idealness parameter $\bar{b}$ and the mass concentration of solid particles in the mixture $K_p$ but decreases by increasing the value of the ratio of the density of solid particles to the initial density of the gas $G_a$ and the gravitation parameter $G_0$. 
\end{description}

\section*{Acknowledgement}
The research of the first author (Ruchi Bajargaan) is supported by CSIR, New Delhi, India vide letter no. 09/045(1264)/2012-EMR-I. The second author (Arvind Patel) thanks to the University of Delhi, Delhi, India for the R\&D grant vide letter no. RC/2015/9677 dated Oct. 15, 2015. 

\section*{Competing Interests}
The authors declare that there is no competing interests regarding the publication of this paper.

\end{document}